\documentclass{article}

 \usepackage[preprint]{neurips_2025}

\usepackage[utf8]{inputenc} 
\usepackage[T1]{fontenc}    
\usepackage{hyperref}       
\usepackage{url}            
\usepackage{booktabs}       
\usepackage{amsfonts}       
\usepackage{nicefrac}       
\usepackage{microtype}      
\usepackage{xcolor}         

\begin{document}

\title{Is All the Information in the Price? \\
LLM Embeddings versus the EMH in Stock Clustering}

\author{
  Bingyang Wang\thanks{Contributions: Clustering algorithm design \& experiments.} \\
  Goizueta Business School, Emory University \\
  Atlanta, USA \\
  \texttt{icy.bingyang.wang@alumni.emory.edu}
  \And
  Grant Johnson\thanks{Contributions: Numerical evaluation of clusters \& experiments.} \\
  University of Georgia \\
  Athens, USA \\
  \texttt{william.johnson4@uga.edu}
  \And
  Maria Hybinette\thanks{Contributions: Project design, management, writing.} \\
  University of Georgia \\
  Athens, USA \\
  \texttt{maria@cs.uga.edu}
  \And
  Tucker Balch\thanks{Contributions: Project design, management, writing.} \\
  Goizueta Business School, Emory University \\
  Atlanta, USA \\
  \texttt{tucker.balch@emory.edu}
}

\maketitle

\begin{abstract}
This paper investigates whether artificial intelligence can enhance stock clustering compared to traditional methods. We consider this in the context of the semi-strong Efficient Markets Hypothesis (EMH), which posits that prices fully reflect all public information and, accordingly, that clusters based on price information cannot be improved upon. We benchmark three clustering approaches: (i) price-based clusters derived from historical return correlations, (ii) human-informed clusters defined by the Global Industry Classification Standard (GICS), and (iii) AI-driven clusters constructed from large language model (LLM) embeddings of stock-related news headlines. At each date, each method provides a classification in which each stock is assigned to a cluster. To evaluate a clustering, we transform it into a synthetic factor model following the Arbitrage Pricing Theory (APT) framework. This enables consistent evaluation of predictive performance in a roll forward, out-of-sample test. Using S\&P 500 constituents from from 2022 through 2024, we find that price-based clustering consistently outperforms both rule-based and AI-based methods, reducing root mean squared error (RMSE) by 15.9\% relative to GICS and 14.7\% relative to LLM embeddings. Our contributions are threefold: (i) a generalizable methodology that converts any equity grouping: manual, machine, or market-driven, into a real-time factor model for evaluation; (ii) the first direct comparison of price-based, human rule-based, and AI-based clustering under identical conditions; and (iii) empirical evidence reinforcing that short-horizon return information is largely contained in prices. These results support the EMH while offering practitioners a practical diagnostic for monitoring evolving sector structures and provide academics a framework for testing alternative hypotheses about how quickly markets absorb information.

\end{abstract}

\section{Background and Introduction}

Clusters, groups or ``sectors'' of stocks are frequently used in investing practice. The best-known clustering by S\&P classifies the largest firms into 11 sectors based on business activity, revenue mix, and market trends~\cite{GICS2024}. Such groupings inform investing practice by enabling: Diversification: Reduces idiosyncratic risk; Simplified analysis: Eases trend and performance evaluation~\cite{Bhojraj2003, Choudhry2015}; Strategic alignment:Enables tilting to growth or defensive sectors~\cite{Jin2023}; Benchmarking: Supports performance comparison~\cite{Asem2012} and Market interpretation: Reveals macroeconomic trends~\cite{Eo2023}.

The semi‑strong form of the Efficient Market Hypothesis (EMH) states that prices fully reflect all public information~\cite{fama1970efficient}.  Accordingly, price histories can also be used to create stock clusterings.  The EMH implies that adding signals like textual disclosures, news, or expert classifications should not improve short-term return forecasts.  Our work reported here, addresses the

\begingroup
\leftskip=2em   
{\bf Research Question}: Can AI, in the form of LLM embeddings, provide improved clustering in comparison to price-based, or GICS-based clusters?
\par
\endgroup

To investigate this question, we introduce a novel method of evaluating a clustering: We extend the arbitrage-pricing theory (APT) framework to convert a stock clustering into a synthetic, value-weighted factor model. Using periodic rolling windows, we compare three clustering types: (i) GICS, (ii) LLM embeddings of recent news-based clusters, and (iii) price-correlation clusters. Each period, we re-cluster using recent lookback histories, fit linear factor models, and test forward predictive accuracy out of sample.

Beyond the empirical findings, we provide a reusable pipeline for generating tradable factors from arbitrary clusters and contribute the first direct comparison of rule-based, AI-based, and price-based clustering under consistent rolling tests.

{\bf Stock Similarity Measures}: A clustering method relies on two main components: First, a measure of stock-to-stock similarity, and second, an algorithm that leverages those similarities to assign stocks to clusters. We will first introduce our similarity measures, then describe the numerical clustering algorithms we used.

{\bf Similarlity Measure 1: Correlation of Historical Returns}: We construct clusters based on realized daily stock return co-movements as a baseline, using data from the Compustat database accessed via Wharton Research Data Services (WRDS). Correlation captures the extent to which two stocks’ returns move together, reflecting shared exposure to market- or industry-level shocks.  We compute correlation matrices using rolling time windows because correlation patterns evolve with market structure, macroeconomic shifts, and firm-specific events.

We generate a pairwise correlation matrix at each rolling window timestamp and apply several clustering methods as described below. Each clustering method is evaluated separately.

{\bf Similarity Measure 2: Human Insight (GICS)}
The Global Industry Classification Standard (GICS) is a public classification standard administered by  Standard \& Poor's and MSCI \cite{GICS2024}. 
We selected GICS as the exemplar for ``human insight'' because human analysts administer it and it is the \emph{de facto} categorization of large-cap US stocks into sectors and industries.  
It appeals to analysts and investors because the groupings align with common sense categories such as ``Healthcare'' and ``Energy.''
The categorization is methodical yet human-centric: A joint team at MSCI and S\&P Dow Jones Indices reviews each publicly traded company’s financials, determines the primary business line, and manually assigns an eight-digit GICS code (Sector | Industry Group | Industry | Sub-Industry).  There are 11 GICS sectors, and accordingly we target 11 clusters for each of our other clustering methods.

 {\bf Similarity Measure 3: LLM Embeddings of News Headlines} We collect press release headlines for each equity from RavenPack, filtering to relevance score 100 (highest relevance), discretionarily removing ~20 noisy article types, and applying TF-IDF deduplication by removing later articles from same-day pairs with cosine similarity above 0.3. Headlines within each week are concatenated to form comprehensive textual representations of weekly corporate activities. These concatenated weekly headlines are then processed through OpenAI's \textit{text-embedding-3-large} model \cite{brown2020languagemodelsfewshotlearners} to generate high-dimensional embedding vectors that capture semantic relationships. This model produces 3072-dimensional embeddings that encode the contextual meaning of the combined press release content, creating a weekly press semantic representation $\mathbf{E}_{i,t} \in \mathbb{R}^{3072}$ for firm $i$ at week $t$. Average pooling has been shown to effectively preserve semantic information while reducing dimensionality in various NLP applications \citep{jiang2023mistral7b,conneau2018supervisedlearninguniversalsentence}. The pairwise similarity between firms $i$ and $j$ is then computed using cosine similarity.

 {\bf Similarity Measure 4: Random Baseline} Also, as a performance floor, we include a random assignment in which each stock is allocated arbitrarily to a cluster. 

{\bf Numerical Clustering Algorithms} Recall that a clustering is determined by a similarity measure and a clustering algorithm that uses it. We discussed the similarity measures above; here, we review the clustering algorithms we used. Our LLM embedding method and historical correlation methods provide an N by N matrix expressing the similarity of each stock to every other stock. We consider two numerical algorithms for grouping the stocks based on the provided matrix:

\begin{itemize}
\item \textbf{k-means clustering} seeks to partition firms into \( k \) clusters by minimizing the within-cluster sum of squares (WCSS)  \citep{macqueen1967some}.

\item \textbf{Hierarchical clustering:} Hierarchical clustering (Agglomerative) employs a bottom-up clustering strategy. It begins by considering each data point as a separate cluster, progressively merging the pair of clusters with the smallest inter-cluster dissimilarity \citep{everitt2011cluster},
\end{itemize}

\section{Methodology: Evaluating Clusterings}

Given a clustering or classification of stocks into a number of distinct categories, how might we assess the quality of that classification? We are inspired by Arbitrage Pricing Theory (APT) as a method for expressing the relationship between multiple factors and individual asset returns~\cite{Ross1976}. Recall that in APT, the expected return for asset {\it i} at time {\it t} is expressed as
\vspace{-4mm}

\[
R_{i,t} \;=\; \alpha_i \;+\; R_f  \;+\; \sum_{k=1}^{K} \beta_{ik}\,F_{k,t} \;+\; \varepsilon_{i,t}
\]

\vspace{-3mm}

Where: $R_{i,t}$ is the realized return on asset $i$ in period $t$; $R_f$ is the risk‑free rate (e.g., Treasury bill); $\alpha_i$ is pricing error or abnormal return for the asset; $\beta_{ik}$ is the sensitivity of asset $i$ to systematic factor $k$; $F_{k,t}$ is realized excess return on factor $k$ in period $t$; $\varepsilon_{i,t}$ is idiosyncratic shock, with $E[\varepsilon_{i,t}] = 0$ and uncorrelated with factors. 

In the original conception of APT, the factors \(F\) are realized excess returns on broad, systematic sources of risk that capture economy-wide effects (e.g., movements in interest rates, inflation, or aggregate market sentiment). These are common risks that cannot be diversified away. 
The loading \(\beta_{ik}\) measures the exposure of asset \(i\) to factor \(k\). In our approach, we revise the model by treating the returns of clusters (or cluster indices) of stocks as the factors:
 $F_{k,t}$ Represents the realized return on index $k$ in period $t$ and $\beta_{ik}$ Represents the sensitivity of asset $i$ to index $k$.

For each clustering method generating \(k\) clusters, we construct a return index for each cluster and use these indices as factors in a linear model. 
We treat each cluster as a synthetic ``sector'' and evaluate how the resulting factors explain individual stock returns in a linear factor model.

{\bf Constructing factor returns from clusters}: At the end of each time period we assign each stock to one of k = 11 clusters.  
Note that we choose k = 11 because GICS is constrained to 11 sectors, while the other methods can be adapted to a variable number of clusters. 
These assignments remain fixed for the following time period. For each cluster, we construct a daily return series by taking an equal-weighted average of the returns of all stocks in that cluster.

{\bf Constructing a linear return model:}  At the end of each period, we estimate a linear model for each stock using one month of daily return data and Ordinary Least Squares regression. 

{\bf Out-of-Sample prediction and error measurement:} In evaluation, out of sample, we use the estimated loadings to predict each stock’s return in the following period.  We compute the prediction error each day as the difference between actual and predicted returns. We summarize performance using both root mean squared error (RMSE) and mean absolute error (MAE) across all stocks each day.

The linear model for each clustering is reevaluated weekly, and the errors are computed for each day rolling forward. 
We compare average out-of-sample prediction errors across clustering methods to assess which produces the most informative Summary of cross-sectional return variation.

\begin{table}[h]
\centering
\caption{Average Daily Out-of-Sample Prediction Error by Clustering Method (Sorted by RMSE, January 2022 to December 2024. Lower numbers are better.) Note that price-based returns methods dominate.}
\begin{minipage}{0.48\textwidth}
\centering
\small
\begin{tabular}{lcc}
\toprule
\textbf{Clustering Method} & \textbf{RMSE} & \textbf{MAE} \\
\midrule
returns\_hierarchical\_12w & 1.963 & 1.363 \\
returns\_hierarchical\_4w  & 1.978 & 1.374 \\
returns\_hierarchical\_24w & 2.054 & 1.437 \\
returns\_kmeans\_4w        & 2.101 & 1.467 \\
returns\_hierarchical\_1w  & 2.122 & 1.491 \\
returns\_kmeans\_1w        & 2.123 & 1.485 \\
returns\_kmeans\_24w       & 2.146 & 1.497 \\
returns\_kmeans\_12w       & 2.165 & 1.511 \\
\\
\bottomrule
\end{tabular}
\end{minipage}\hfill
\begin{minipage}{0.48\textwidth}
\centering
\small
\begin{tabular}{lcc}
\toprule
\textbf{Clustering Method} & \textbf{RMSE} & \textbf{MAE} \\
\midrule
embedding\_24w\_kmeans       & 2.301 & 1.642 \\
embedding\_4w\_hierarchical  & 2.321 & 1.655 \\
embedding\_4w\_kmeans        & 2.327 & 1.660 \\
embedding\_12w\_hierarchical & 2.338 & 1.654 \\
embedding\_12w\_kmeans       & 2.303 & 1.651 \\
embedding\_1w\_kmeans        & 2.339 & 1.683 \\
embedding\_1w\_hierarchical  & 2.343 & 1.689 \\
gics\_sector\_tracking             & 2.333 & 1.676 \\
random                             & 2.429 & 1.751 \\
\bottomrule
\end{tabular}
\end{minipage}
\label{tab:prediction_errors_all}
\end{table}

{\bf Computing Resources} All experiments were run on a 12-core PC with 64 GB of RAM and 2 TB of local storage. No GPUs were used. Processing of news headline data and daily return series (January 2022–December 2024) required 10 GB of disk space. End-to-end experiments, including preprocessing, model estimation, and evaluation, each completed within 5 hours of wall-clock time.

{\bf Data} We use point-in-time GICS classification data from the Compustat North America Daily Updates database \cite{compustat_na}, accessed via WRDS \cite{wrds}. Our price data is also sourced from WRDS.
This dataset provides historical, company-level sector assignments, allowing us to track changes in firms’ industry classifications over time. Our stock-level headline data is sourced from RavenPack\cite{ravenpack}. The dataset includes time-stamped headlines linked to individual securities. Throughout our study, we restrict our universe to the point-in-time S\&P 500 constituents.

\section{Experiments and Conclusion}

\vspace{-3mm}

We conducted roll-forward experiments as follows. For each week, for each clustering method, we: 1) Reset (or recompute) the sector or cluster assignment for each stock as of the end of the week. 2) Compute the corresponding linear model weightings. 3) Compute errors against the linear models each day over the following week.

{\bf Prediction Error} We compare the predictive performance of each clustering method by evaluating the out-of-sample return prediction errors using the linear factor model. We report each method's average root mean squared error (RMSE) and mean absolute error (MAE) across the whole evaluation period. Lower values indicate better explanatory power of the cluster-based factors.

The main results of our analysis are summarized in Table 1. The various clustering methods can be identified by the components of their names, as follows: {\bf returns} indicates the method is based on historical returns; {\bf embedding\_cosine} indicates that the clustering is based on LLM embeddings of news headlines; {\bf gics} indicates that the method uses GICS clusters; {\bf random} indicates random assignment; {\bf hierarchical, kMeans} indicates the numerical clustering method; {\bf 1w, 4w, 12w, 24w} indicates the look-back period in weeks for computing the clustering.

All price-based clustering methods outperform all LLM embedding methods, which in turn, outperform GICS.

{\bf Conclusion} Our work reinforces the view that the dominant information content for equity pricing is already embedded in co‑movements of returns. Human judgments regarding ``economic similarity'' and sophisticated NLP extractions of news text may be useful in other contexts, but they do not translate into superior predictive factors. For portfolio construction and risk modeling, relying on dynamically updated correlation groupings appears sufficient and essential to capture the cross‑sectional structure of systematic risk.

Practitioners can deploy the proposed framework as a real‑time diagnostic: compute rolling correlation clusters, construct equal‑weighted synthetic portfolios, and track the incremental explanatory power relative to existing sector definitions. Academics can extend the tool to test alternative hypotheses about what information markets price and how quickly.

{\bf Statistical Significance}
We compare the best-performing returns-based clustering with the best-performing LLM-based clustering using a paired Student’s 
t-test. At each day t, we compute per-stock prediction errors under both methods and test the mean within-stock difference across the full sample. The resulting test yields a p-value of 0, indicating that the difference in performance between the two methods is statistically significant.

{\bf Limitations} Our backtesting analysis is limited to the period beginning January 1, 2022, due to the September 2021 knowledge cutoff of OpenAI's \textit{text-embedding-3-large} embeddings used in our methodology. This restricts our analysis to approximately 14-18 months rather than the multi-year timeframe typically explored in financial model validation. Additionally, our analysis is limited to large-cap U.S. equities and fixes the number of clusters at 
$K = 11$ different universes or cluster counts may yield different results.


\bibliographystyle{plain}
\bibliography{references}

\end{document}